\documentclass[12pt,english]{article}
\usepackage[T1]{fontenc}
\usepackage[latin9]{inputenc}
\usepackage[letterpaper]{geometry}
\usepackage{textcomp}
\usepackage{amsthm}
\usepackage{amsmath}
\usepackage{graphicx}
\usepackage{setspace}
\usepackage{amssymb}
\usepackage[authoryear]{natbib}
\makeatletter

\providecommand{\tabularnewline}{\\}

\newcommand{\lyxaddress}[1]{
\par {\raggedright #1
\vspace{1.4em}
\noindent\par}
}
\theoremstyle{plain}
\newtheorem{thm}{Theorem}
 \theoremstyle{definition}
  \newtheorem{example}[thm]{Example}
  \theoremstyle{remark}

\DeclareMathOperator{\N}{N}

\DeclareMathOperator{\lfdr}{LFDR}

\makeatother

\usepackage{babel}

\begin{document}

\title{Large-scale interval and point estimates from an empirical Bayes
extension of confidence posteriors}

\author{David R. Bickel}

\maketitle

\lyxaddress{Ottawa Institute of Systems Biology\\
Department of Biochemistry, Microbiology, and Immunology\\
Department of Mathematics and Statistics \\
University of Ottawa\\
451 Smyth Road\\
Ottawa, Ontario, K1H 8M5}
\begin{abstract}
In statistical genomics, bioinformatics, and neuroinformatics, truth
values of multiple hypotheses are often modeled as random quantities
of a common mixture distribution in order to estimate false discovery
rates (FDRs) and local FDRs (LFDRs). Unfortunately, the FDRs or LFDRs are typically reported with conventional
confidence intervals or point estimates of a parameter of interest
rather than shrunken interval and point estimates consonant with the
hierarchical model that underlies LFDR estimation. In a pure Bayesian
approach, the shrunken estimates may be derived from a fully specified
prior on the parameter of interest. While such a prior may in principle
be estimated under the empirical Bayes framework, published methods
taking that approach require strong parametric assumptions about the
parameter distribution. 

The proposed approach instead extends the confidence posterior distribution
to the semi-parametric empirical Bayes setting. Whereas the Bayesian
posterior is defined in terms of a prior distribution conditional
on the observed data, the confidence posterior is defined such that
the probability that the parameter value lies in any fixed subset
of parameter space, given the observed data, is equal to the coverage
rate of the corresponding confidence interval. A confidence posterior
that has correct frequentist coverage at each fixed parameter value
is combined with the estimated LFDR to yield a parameter distribution
from which interval and point estimates are derived within the framework
of minimizing expected loss. The point estimates exhibit suitable
shrinkage toward the null hypothesis value, making them practical
for automatically ranking features in order of priority. The corresponding
confidence intervals are also shrunken and tend to be much shorter
than their fixed-parameter counterparts, as illustrated with gene
expression data. Further, simulations confirm a theoretical argument
that the shrunken confidence intervals cover the parameter at a higher-than-nominal
frequency.
\end{abstract}

\newpage{}

\section{\label{sec:Introduction}Introduction}

By enabling simultaneous tests of whether each of thousands of genes
represented on a microarray is differentially expressed across experimental
or clinical conditions, advances in biotechnology have lead to increased
use of the false discovery rate (FDR) as a solution to extreme multiple
comparisons problems. As a result, the statistical community has developed
more general and more powerful methods of controlling what \citet{RefWorks:288}
called the FDR while proposing new definitions of the FDR \citep{Farcomeni2008347}.
The alternative strategy of estimating rather than controlling the
FDR in turn led \citet{RefWorks:53} to propose estimating the local
false discovery rate (LFDR), a limiting case of an FDR. Recently,
\citet{mse2008} found LFDR estimators to perform well in terms of
prediction error computed with gene expression microarray data, and
\citet{Schwartzman200971} applied LFDR methods to the analysis of
neuroimaging data. (The terminology here follows the empirical Bayes
convention of referring to predictors of random quantities such as
the LFDR as estimators.)

FDR estimation begins with the reduction of the data directly bearing
on each null hypothesis to a low-dimensional statistic such as a Student
\emph{t} statistic or a p-value and the specification of a subset
of reduced-data space called the \emph{rejection region}. In a general
empirical Bayes framework, the \emph{Bayesian FDR }(BFDR) is the conditional
probability that a null hypothesis is true given that it is rejected,
that is, given that its statistic lies in the rejection region \citep{RefWorks:54}.
Relaxing the requirement that all null hypotheses share the same rejection
region and instead setting the rejection region of each null hypothesis
to the set containing only the observed value of its statistic generates
a different BFDR for each hypothesis; such a BFDR is called an LFDR.
The LFDR of a null hypothesis is the conditional probability that
it is true given that its statistic is equal to its observed value.
Thus, estimates of the LFDR are often interpreted as approximations
of fully Bayesian posterior probabilities that could have been computed
were a suitable joint prior distribution of all unknown parameters
available. 

However, from a hierarchical Bayesian perspective, the LFDR estimate
suffers as an approximation of a hypothesis posterior probability
in its failure to incorporate the uncertainty in the parameters. Similarly,
from a frequentist perspective, the point estimate of the LFDR would
seem less desirable than an interval estimate of the LFDR since the
latter would reflect uncertainty in the true value of the LFDR, and
correlations between data of different biological features can introduce
substantial variability into FDR and LFDR estimates \citep{antiFDRcontrol2004,Qiu2005i}.
\citet{efron_correlated_2010} addressed the problem of estimate accuracy
by providing asymptotic bounds on the confidence limits of the FDR
in the presence of correlation between statistics. Nonetheless, it
is not clear how reporting a standard error or confidence interval
for the LFDR of each of thousands of null hypotheses would facilitate
the interpretation of the results \citep{Westfall_efron_correlated_2010}. 

Fortuitously, as the probability of hypothesis truth, the LFDR itself
is of much less direct biological interest than is the random parameter
about which a hypothesis is formulated. Both the Bayesian and frequentist
criticisms that LFDR estimation inadequately incorporates uncertainty
in the parameter distribution may be answered by constructing conservative
confidence intervals for the random parameters of interest under the
finite mixture model that underlies LFDR estimation, as \citet{RefWorks:1273}
accomplished for a mixture of two normal distributions. 

The assumption of a known parametric model for the random parameter
will be dropped in Section \ref{sec:Frequentist-posteriors}, which
instead uses a \emph{confidence posterior}, a continuous distribution
of confidence levels for a given hypothesis on the basis of nested
confidence intervals. Like the Bayesian posterior, the confidence
posterior is an inferential (non-physical) distribution of the parameter
of interest that is coherent according to various decision theories
\citep{CoherentFrequentism,conditional2009}. Unlike the Bayesian
posterior, the confidence posterior does not require specification
of or even compatibility with any prior distribution. The interest
parameter $\theta$ is a subparameter of the full parameter $\xi$,
which specifies the sampling probability distribution $P_{\xi}$.
In the case of a one-dimensional parameter of interest, the confidence
posterior is completely specified by a set of nested confidence intervals
with exact coverage rates. Given the observed realization $x$ of
a $P_{\xi}$-distributed data vector $X$, the confidence posterior
distribution $P^{x}$ is defined such that the probability that the
parameter lies in a given interval $\left[\theta^{\prime},\theta^{\prime\prime}\right]$
is equal to the coverage rate of the confidence interval equal to
that given interval. That is, \begin{equation}
P^{x}\left(\vartheta\in\left[\theta^{\prime},\theta^{\prime\prime}\right]\right)=P_{\xi}\left(\theta\in\Theta_{\rho}\left(X\right)\right)=\rho,\label{eq:coverage}\end{equation}
where $\vartheta$ is the random interest parameter of distribution
$P^{x}$, and $\Theta_{\rho}$ is the interval estimator with rate
$\rho$ of coverage constrained such that $\Theta_{\rho}\left(x\right)=\left[\theta^{\prime},\theta^{\prime\prime}\right]$.
To distinguish $P^{x}\left(\vartheta\in\left[\theta^{\prime},\theta^{\prime\prime}\right]\right)$
for a specified hypothesis that $\theta\in\left[\theta^{\prime},\theta^{\prime\prime}\right]$
from a confidence interval of a specified confidence level $\rho$,
\citet{Polansky2007b} called the former an \emph{observed confidence
level}. Marginalizing the confidence posterior over the estimated
LFDR as the probability of null hypothesis truth shrinks the confidence
posterior toward the parameter value of the null hypothesis. Shrunken
interval and point estimates are then derived from the marginal confidence
posterior.

The use of the shrunken estimates will be illustrated in Section \ref{sec:Case-study}
with an application to gene expression data. Section \ref{sec:Simulation-study}
reports a simulation study of the shrunken confidence interval and
point estimates. Section \ref{sec:Discussion} closes with a summary
of the findings.

\section{\label{sec:Frequentist-posteriors}Frequentist posteriors for shrunken
estimates}

\subsection{\label{sub:Confidence-posterior-distributions}Confidence posterior
distributions}

Considering the observed data vector $x\in\mathcal{X}^{n}$ as a sample
from a distribution in the parametric family $\left\{ P_{\xi}:\xi\in\Xi\right\} $
parameterized by $\xi$ in $\Xi\subseteq\mathbb{R}^{d}$, the value
in $\Theta\subseteq\mathbb{R}^{1}$ of the parameter of interest is
denoted by $\theta\left(\xi\right)$. The function $F_{\bullet}:\mathcal{X}^{n}\times\Theta\rightarrow\left[0,1\right]$
is called a \emph{significance function} if $F_{X}\left(\theta\right)$
is distributed uniformly between 0 and 1 for all $\theta\in\Theta$
and if $F_{x}$ is a cumulative distribution function for all $x\in\mathcal{X}^{n}$.
Due to the latter property, the significance function evaluated at
$x$ is also known as the \emph{confidence distribution} \citep{RefWorks:60,RefWorks:130},
but \citet{Efron19933} and \citet{RefWorks:127} used that term in
the sense of the following probability distribution. Given any $x\in\mathcal{X}^{n}$,
the \emph{confidence posterior} $P^{x}$ is the probability measure
on measurable space $\left(\Theta,\mathcal{B}\left(\Theta\right)\right)$
of a random quantity $\vartheta$ such that $P^{x}\left(\vartheta\le\theta\right)=F_{x}\left(\theta\right)$
for all $\theta\in\Theta$, where each $\mathcal{B}\left(\Theta\right)$
is the Borel $\sigma$-field on $\Theta$. It is easy to verify that
equation (\ref{eq:coverage}) holds for all $\xi\in\Xi$ and $\theta^{\prime},\theta^{\prime\prime}\in\Theta$
and for any $\mathcal{X}^{n}$-measurable function $\Theta_{1-\alpha_{1}-\alpha_{2}}$
that satisfies \[
\Theta_{1-\alpha_{1}-\alpha_{2}}\left(x\right)=\left[F_{x}^{-1}\left(\alpha_{1}\right),F_{x}^{-1}\left(1-\alpha_{2}\right)\right]\]
for every $x\in\mathcal{X}^{n}$ and every $\alpha_{1}$ and $\alpha_{2}$
in $\left[0,1\right]$ such that $\alpha_{1}+\alpha_{2}<1$. 

As a Kolmogorov probability measure on parameter space, the confidence
posterior yields coherent decisions in the sense of minimizing expected
loss, as does the Bayesian posterior, and yet without dependence on
any prior distribution \citep{CoherentFrequentism,conditional2009}.
For example, the confidence posterior mean, minimizing expected squared
error loss, is $\bar{\vartheta}_{x}=\int_{\Theta}\vartheta dP^{x}\left(\vartheta\right)$,
and the confidence posterior $p$-quantile, minimizing expected loss
for a threshold-based function of $p$ \citep[App. B]{CarlinLouis3},
is $\vartheta\left(p\right)$ such that $p=P^{x}\left(\vartheta<\vartheta\left(p\right)\right).$\begin{example}
\label{exa:paired-1gene}Assume that $Y_{j},$ the observable, log-transformed
difference in levels of expression of a particular gene between the
$j$th individual of the treatment group and the $j$th individual
of the control group, is a normally distributed random variable of
unknown mean $\theta$ and unknown\emph{ }variance $\sigma^{2}.$
For the observed differences $y_{1},\dots,y_{n}\in\mathcal{X}=\mathbb{R}$,
the $n$-tuple $x=\left\langle y_{1},\dots,y_{n}\right\rangle $ is
thus modeled as a realization of $X=\left\langle Y_{1},\dots,Y_{n}\right\rangle ,$
with $Y_{i}$ independent of $Y_{j}$ for all $i\ne j.$ Then the
one-sample $t$-statistic $\tau\left(X\right)$ has the Student \emph{t}
probability distribution of $n-1$ degrees of freedom. The significance
function $F_{\bullet}$ and confidence posterior $P^{x}$ satisfy\begin{eqnarray}
F_{x}\left(\theta\right)=P^{x}\left(\vartheta\le\theta\right) & = & P_{\left\langle \theta,\sigma\right\rangle }\left(\tau\left(X\right)\ge\tau\left(x\right)\right)\label{eq:t-posterior}\end{eqnarray}
for all $\theta\in\mathbb{R}.$ 
\end{example}

Model $x_{i}\in\mathcal{X}^{n}$, the $i$th of $m$ observed data
vectors each corresponding to a gene or other biological feature,
as a sample of $P_{\xi_{i}}$ with $\xi_{i}\in\Xi$ as the value of
the full parameter and $\theta_{i}=\theta\left(\xi_{i}\right)$ as
the value of the interest parameter. The $i$th null hypothesis asserts
that $\theta_{i}=\theta_{0}$, where $\theta_{0}$ may be any specified
value in $\Theta$.

\subsection{Empirical Bayes}

Empirical Bayes estimators of the LFDR flow from variations of the
following hierarchical mixture model of a data set that has been reduced
to a single scalar statistic per null hypothesis. Examples of such
statistics include test statistics, \emph{p}-values, and, as in \citet{RefWorks:55},
probit transformations of \emph{p}-values. With an $\mathcal{X}^{n}$-measurable
map $\tau:\mathcal{X}^{n}\rightarrow\mathcal{T}$, the observed statistic
$t_{i}=\tau\left(x_{i}\right)$ associated with the null hypothesis
that $\theta_{i}=\theta_{0}$ is assumed to be a realization of the
random statistic $T_{i}$ of the two-component mixture probability
density function $f$ such that \begin{equation}
f\left(t\right)=\pi_{0}f_{0}\left(t\right)+\pi_{1}f_{1}\left(t\right)\label{eq:mixture}\end{equation}
for all $t\in\mathcal{T}$, where $\pi_{0}\in\left[0,1\right],$ $\pi_{1}=\pi_{0}-1,$
and $f_{0}$ and $f_{1}$ are probability density functions (PDFs)
corresponding to the null and alternative hypotheses, respectively.
As the unknown PDF of the statistic conditional on the alternative
hypotheses, $f_{1}$ is estimated by some $\hat{f}_{1}$. Herein,
$f_{0}$ is considered the known PDF of the statistic conditional
on the null hypothesis, but it can instead be estimated if $m$ is
sufficiently large \citep{RefWorks:55}. The mixture distribution
can be equivalently specified by $f_{A}$, where $A$ is a random
quantity equal to 0 with probability $\pi_{0}$ and to 1 with probability
$\pi_{1}$. 

Let $\mathbf{t}=\left\langle t_{1},\dots,t_{m}\right\rangle $ and
$\mathbf{T}=\left\langle T_{1},\dots,T_{m}\right\rangle $. (Since
$T_{i}$ and $T_{j}$ are identically distributed for all $i,j\in\left\{ 1,\dots,m\right\} $
under the mixture model \eqref{eq:mixture}, the model of Section
\ref{sub:Confidence-posterior-distributions} obtains conditionally
for the random $\theta_{i}$.) The local false discovery rate for
the $i$th statistic is defined as the posterior probability that
the $i$th null hypothesis is true:\[
\ell_{i}=\lfdr\left(t_{i}\right)=P\left(A_{i}=0\vert T_{i}=t_{i}\right)=\frac{\pi_{0}f_{0}\left(t_{i}\right)}{f\left(t_{i}\right)}.\]
It is estimated by replacing $\pi_{0}$ and $f_{1}$ with their estimates:\[
\hat{\ell}_{i}=\frac{\hat{\pi}_{0}f_{0}\left(t_{i}\right)}{\hat{\pi}_{0}f_{0}\left(t\right)+\left(1-\hat{\pi}_{0}\right)\hat{f}_{1}\left(t\right)}.\]

\subsection{\label{sub:Marginal-posterior-probabilities}Extended confidence
posteriors}

Marginalization over hypothesis truth leads to estimated posterior
probabilities that each parameter of interest is less than, equal
to, and greater than the parameter value of the null hypothesis. Such
probabilities are coherent with each confidence posterior given the
truth of the alternative hypothesis according to the confidence-based
decision theory of \citet{CoherentFrequentism} and \citet{conditional2009}. 

Consider the probability distribution $P^{\left(i\right)}$ of which
each $P^{x_{i}}$ is a conditional probability distribution of $\vartheta_{i}$
given $\theta_{i}\ne\theta_{0}$, of which $\delta_{\theta_{0}}$,
the Dirac measure at $\theta_{0}$, is a conditional probability distribution
of $\vartheta_{i}$ given $\theta_{i}=\theta_{0}$, and according
to which $\ell_{i}$ is the probability that $\theta_{i}=\theta_{0}$.
That is, $P^{\left(i\right)}\left(A_{i}=0\right)=\ell_{i}$ and, for
all $\theta\in\Theta$, \[
P^{\left(i\right)}\left(\vartheta_{i}\le\theta\vert A_{i}=1\right)=P^{x_{i}}\left(\vartheta_{i}\le\theta\right)\]
and, with the function $1_{S}\left(\bullet\right)$ respectively indicating
membership and non-membership in $S$ by 1 and 0, \[
P^{\left(i\right)}\left(\vartheta_{i}\le\theta\vert A_{i}=0\right)=\delta_{\theta_{0}}\left(\vartheta_{i}\le\theta\right)=1_{\left[\theta_{0},\infty\right)}\left(\theta\right).\]
In the more succinct mixture notation,\begin{equation}
\vartheta_{i}\sim P^{\left(i\right)}=\ell_{i}\delta_{\theta_{0}}+\left(1-\ell_{i}\right)P^{x_{i}}.\label{eq:marginal-target}\end{equation}
 Since $P^{\left(i\right)}$ as the inferential parameter distribution
follows from applying Kolmogorov probability theory to the base distributions
$\pi_{\bullet}$, $\delta_{\theta_{0}}$, and $P^{x_{i}}$, decisions
made on its basis are those that would be required by the base distributions
in the framework of minimizing expected loss with respect to a confidence
posterior distribution \citep{CoherentFrequentism,conditional2009}
and, more generally, with respect to any parameter distribution \citep[e.g.,][]{MaxUtility1944,RefWorks:126}.
For example, as the posterior median $F_{x_{i}}^{-1}\left(1/2\right)$
minimizes the expected absolute loss involved in estimating $\theta_{i}$
conditional on $A_{i}=1$, the median $\bar{\vartheta}_{i}$ of $P^{\left(i\right)}$
does so marginally. Thus, $P^{\left(i\right)}$ will be called the
\emph{marginal confidence posterior} and $P^{x_{i}}$ the \emph{conditional
confidence posterior} given the truth of the alternative hypothesis.
Adapting the terminology of \citet{Polansky2007b} concerning fixed
parameters of interest, $P^{\left(i\right)}$-probabilities and $P^{x_{i}}$-probabilities
of hypotheses will be called \emph{(observed) marginal }and \emph{conditional
confidence levels}, respectively.

Since $\pi_{\bullet}$ is unknown, the marginal confidence posterior
will be estimated by\begin{equation}
\hat{P}^{\left(i\right)}=\hat{\ell}_{i}\delta_{\theta_{0}}+\left(1-\hat{\ell}_{i}\right)P^{x_{i}},\label{eq:marginal-estimate}\end{equation}
which resembles the marginal empirical Bayes posterior\emph{ }from
which \citet{RefWorks:1273} derived conservative confidence intervals
under a parametric model. (\citet{RefWorks:1275} similarly derived
empirical Bayes interval estimates conditional on \textbf{$A_{i}=1$}
in order to contrast them with estimates that control a false coverage
rate \citep{RefWorks:1276}.) The two posterior distributions differ
in that $P^{x_{i}}$ is a confidence posterior rather than the Bayesian
posterior $P_{\text{prior}}\left(\bullet\vert A_{i}=1,T_{i}=t_{i}\right)$,
which requires specification or estimation of $P_{\text{prior}}\left(\bullet\vert A_{i}=1\right)$,
a prior distribution of $\theta_{i}$ conditional on the truth of
the alternative hypothesis. For ease of reading, $\hat{P}^{\left(i\right)}$-probabilities
of hypotheses will be called \emph{(observed) marginal confidence
levels }even though they are more precisely estimates of such levels.
\begin{example}
\label{exa:paired-multiple-genes}Generalizing Example \ref{exa:paired-1gene}
to multiple genes, let $x_{i}$ denote the $n$-tuple of log-transformed
differences in levels of expression of the $i$th of $m$ genes. The
$i$th null hypothesis is that the $i$th gene is equivalently expressed
$\left(\theta_{i}=0\right)$ as opposed to differentially expressed
$\left(\theta_{i}\ne0\right)$. Further, let $P^{x_{i}}$ denote the
corresponding confidence posterior defined by equation \eqref{eq:t-posterior}
and the normality and conditional independence assumptions of Example
\ref{exa:paired-1gene}. That $P^{x_{i}}$ is mathematically equivalent
to the Bayesian posterior $P_{\text{prior}}\left(\bullet\vert A_{i}=1,T_{i}=t_{i}\right)$
formulated by the uniform {}``distribution'' (Lebesgue measure)
as the prior for $\theta_{i}$ and integrating over the standard deviation
$\sigma$ with respect to the posterior from the independent prior
density proportional to $1/\sigma$. Since the prior is not a Kolmogorov
probability distribution, the estimated posterior odds given by multiplying
the estimated prior odds $\left(1-\hat{\pi}_{0}\right)/\hat{\pi}_{0}$
by the Bayes factor is undefined \citep{RefWorks:1023,mse2008}.
Thus, there is no prior distribution that corresponds to $\hat{P}^{\left(i\right)}$
in this example. (\citet{mse2008} used a predictive distribution
on the basis of an intuitively motivated posterior equivalent to $\hat{P}^{\left(i\right)}$
to assess the performance of various predictors of gene expression
data.)
\end{example}

\subsection{\label{sub:Estimates}Point and interval estimates}

Were the marginal confidence posterior $P^{\left(i\right)}$ known,
its mean and median would respectively minimize expected square-error
and absolute loss incurred by estimating $\theta_{i}$ (§\ref{sub:Confidence-posterior-distributions}),
and the odds for betting that $\theta_{i}$ lies in some subset $\Theta^{\prime}$
of $\Theta$ would be $P^{\left(i\right)}\left(\vartheta_{i}\in\Theta^{\prime}\right)/P^{\left(i\right)}\left(\vartheta_{i}\in\Theta\backslash\Theta^{\prime}\right)$,
a ratio of two observed marginal confidence levels \citep{CoherentFrequentism}. 

Those decision-theoretic considerations suggest estimating $\theta_{i}$
by the mean or median of $\hat{P}^{\left(i\right)}$ and constructing
the $\left(1-\alpha_{1}-\alpha_{2}\right)100\%$ \emph{marginal confidence
interval} $\left[\hat{F}_{\left(i\right)}^{-1}\left(\alpha_{1}\right),\hat{F}_{\left(i\right)}^{-1}\left(1-\alpha_{2}\right)\right]$
from the \emph{marginal significance function} $\hat{F}_{\left(i\right)}$
defined by $\hat{F}_{\left(i\right)}\left(\theta\right)=\hat{P}^{\left(i\right)}\left(\vartheta_{i}\le\theta\right)$
for all $\theta\in\Theta$. Inverting $\hat{F}_{\left(i\right)}$
gives, for any $\alpha\in\left[0,1\right]$, \begin{equation}
\hat{F}_{\left(i\right)}^{-1}\left(\alpha\right)=\begin{cases}
F_{x_{i}}^{-1}\left(\alpha/\left(1-\hat{\ell}_{i}\right)\right) & \text{if }F_{x_{i}}^{-1}\left(\alpha/\left(1-\hat{\ell}_{i}\right)\right)<\theta_{0}\\
F_{x_{i}}^{-1}\left(1-\left(1-\alpha\right)/\left(1-\hat{\ell}_{i}\right)\right) & \text{if }F_{x_{i}}^{-1}\left(1-\left(1-\alpha\right)/\left(1-\hat{\ell}_{i}\right)\right)>\theta_{0}\\
\theta_{0} & \text{otherwise}\end{cases},\label{eq:shrunken-CI}\end{equation}
where $F_{x_{i}}$ is the \emph{conditional significance function
}defined in Section \ref{sub:Confidence-posterior-distributions}. 

While the $\hat{P}^{\left(i\right)}$-probability that $\vartheta_{i}$
lies in the interval estimate is exactly $\left(1-\alpha_{1}-\alpha_{2}\right)100\%$
by construction, it does not have exact frequentist coverage. However,
two limiting cases suggest that the marginal confidence interval covers
the random value of $\theta_{i}$ at a relative frequency greater
than the nominal rate $\rho=\left(1-\alpha_{1}-\alpha_{2}\right)100\%$:\begin{equation}
\lim_{\lambda\rightarrow0}P_{\text{true}}\left(\theta_{i}\le\hat{F}_{\left(i\right)}^{-1}\left(\alpha\right)\vert A_{i}=1,\hat{\ell}_{i}\le\lambda\right)=P_{\text{true}}\left(F_{x_{i}}\left(\theta_{i}\right)\le\alpha\vert A_{i}=1\right)=\alpha\label{eq:approaching0}\end{equation}

\begin{equation}
\lim_{\lambda\rightarrow1}P_{\text{true}}\left(\theta_{i}\le\hat{F}_{\left(i\right)}^{-1}\left(\alpha\right)\vert A_{i}=0,\hat{\ell}_{i}\ge\lambda\right)=P_{\text{true}}\left(\theta_{i}=\theta_{0}\vert A_{i}=0\right)=1,\label{eq:approaching1}\end{equation}
where $0<\alpha<1$ and $0<\lambda<1$; $P_{\text{true}}$ is the
sampling distribution of $\left\langle \theta_{i},X_{i}\right\rangle $.
To the extent that $1-\pi_{0}$ is small, the actual coverage rate
$P_{\text{true}}\left(\theta_{i}\in\left[\hat{F}_{\left(i\right)}^{-1}\left(\alpha_{1}\right),\hat{F}_{\left(i\right)}^{-1}\left(1-\alpha_{2}\right)\right]\right)$
is dominated by the rate conditional on $A_{i}=0$. For that reason,
equation \eqref{eq:approaching1} indicates that, inasmuch as $\hat{\pi}_{0}$
is a positively biased estimator of some sufficiently small $\pi_{0}$,
the confidence intervals derived from $\hat{P}^{\left(i\right)}$
are conservative in the sense that they include the random value
of $\theta_{i}$ at a relative frequency higher than the nominal $\left(1-\alpha_{1}-\alpha_{2}\right)100\%$
level for any $\alpha_{1},\alpha_{2}\in\left[0,1\right]$ such that
$\alpha_{1}+\alpha_{2}<1$. 

Likewise, the $\hat{P}^{\left(i\right)}$-posterior median $\hat{F}_{\left(i\right)}^{-1}\left(1/2\right)$,
corresponding to the degenerate $0\%$ confidence interval $\left[\hat{F}_{\left(i\right)}^{-1}\left(50\%\right),\hat{F}_{\left(i\right)}^{-1}\left(50\%\right)\right]$,
is conservative in the sense that a positive bias in $\hat{\ell}_{i}$
pulls the estimate $\hat{F}_{\left(i\right)}^{-1}\left(1/2\right)$
toward $\theta_{0}$. The extent of the conservatism of both point
and interval estimates was quantified by simulation as described in
Section \ref{sec:Simulation-study}.

\section{\label{sec:Case-study}Application to gene expression}

Microarray technology enables measurement of the expression levels
of thousands of genes for each \emph{biological} \emph{replicate},
an organism\emph{ }or set of organisms studied. Most microarray experiments
are designed to determine which genes to consider differentially expressed
across two conditions, conveniently called \emph{treatment} and \emph{control}.
Investigators initially relied on estimates of an average ratio of
expression under the treatment condition to that under the control
condition without using hypothesis tests. As statisticians have responded
with extensive research on multiple comparison procedures, biologists
have moved to ignoring estimated levels of differential expression
for all genes that do not correspond to rejected null hypotheses. 

In response, \citet{shrink2009} proposed the prioritization of genes
for further study by shrunken estimates of differential expression
levels, much as \citet{ISI:000253401400024} and \citet{ISI:000280145100013}
suggested prioritizing single-nucleotide polymorphisms (SNPs) by shrunken
estimates of odds ratios. Whereas \citet{shrink2009}, following \citet{RefWorks:26}
and \citet{mse2008}, used a heuristic estimate equal in value to
the posterior mean with respect to $\hat{P}^{\left(i\right)}$, the
posterior median has the advantage of invariance to reparameterization.
Since, in addition, the posterior median is a limiting case of a confidence
interval (§\ref{sub:Estimates}), it will be used as the point estimate
alongside the interval estimate.

While point estimation is practical for ranking genes in order of
priority, interval estimates are needed to quantify their reliability.
In place of the commonly used confidence intervals that do not account
for multiple comparisons, we will report the shrunken confidence intervals
of equation \eqref{eq:shrunken-CI}.

The amount the gene expression differs between mutant tomatoes and
wild type (WT) tomatoes were estimated for $n=6$ mutant-WT ({}``treatment-control'')
pairs at 3 days after the breaker stage of ripening; the microarrays
represent 13,440 genes. \citet{RefWorks:8} provide details of the
fruit development experiments conducted. Due to the pairing of mutant
and WT biological replicates, the normal model and confidence posteriors
of Examples \ref{exa:paired-1gene} and \ref{exa:paired-multiple-genes}
were used. Each local false discovery rate was estimated by the {}``theoretical
null'' method of \citet{RefWorks:208} since simulations indicate
that the {}``empirical null'' method applied to the model of \ref{exa:paired-1gene}
and \ref{exa:paired-multiple-genes} loses power in the presence of
heavy-tailed data like that of gene expression \citep{conditional2009}.

\begin{figure}[ph]
\includegraphics{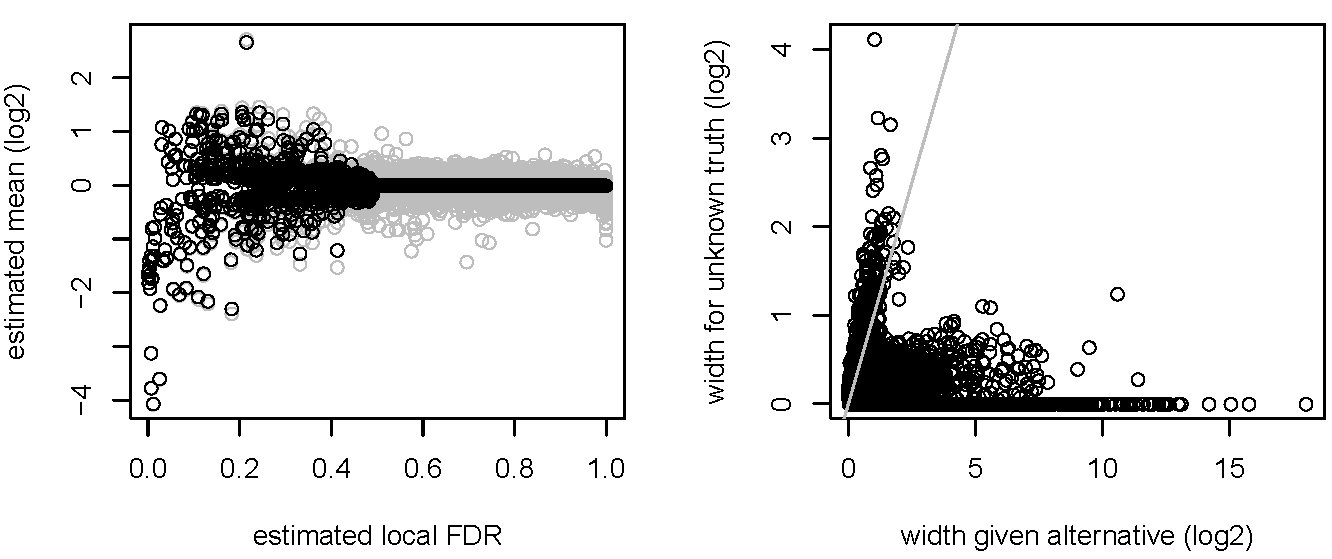}

\caption{Point and interval estimates of gene expression.\label{fig:expression}}

\end{figure}

Each circle of Fig. \ref{fig:expression} represents a point or interval
estimate of $\theta_{i}$ for a gene. The left-hand side displays
the posterior median of $\vartheta_{i}$ versus $\hat{\ell}_{i}$
on the basis of each marginal confidence posterior $\hat{P}^{\left(i\right)}$
(black) and each conditional confidence posterior $P^{x_{i}}$ (gray).
Stronger shrinkage is evident at higher values of $\hat{\ell}_{i}$.

The right-hand side of Fig. \ref{fig:expression} features the width
of the confidence interval from each $\hat{P}^{\left(i\right)}$ versus
the width of the confidence interval from each $P^{x_{i}}$. It is
apparent that the use of the marginal confidence posterior in place
of the conditional confidence posterior tends to substantially reduce
interval width. 

\begin{figure}
\includegraphics{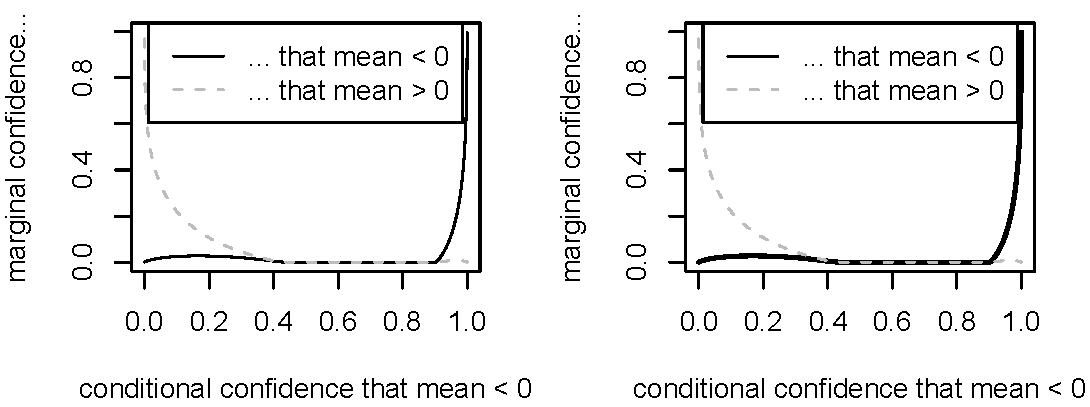}

\caption{Observed confidence levels about mean differential expression.\label{fig:Observed-confidence-levels}}

\end{figure}

In Fig. \ref{fig:Observed-confidence-levels}, observed marginal confidence
levels are plotted against observed conditional confidence levels
to show how much inferential probability each attributes to the hypothesis
that $\theta_{i}<0$ and to the hypothesis that $\theta_{i}>0$. The
horizontal axis has $P^{x_{i}}\left(\vartheta_{i}<0\right)$, which
is equal to $P^{\left(i\right)}\left(\vartheta_{i}<0\vert A_{i}=1\right)$
and to $1-P^{x_{i}}\left(\vartheta_{i}>0\right)$. The vertical axis
has $\hat{P}^{\left(i\right)}\left(\vartheta_{i}<0\right)$ in black
and $\hat{P}^{\left(i\right)}\left(\vartheta_{i}>0\right)$ in gray;
these marginal confidence levels do not total 100\% since $\hat{P}^{\left(i\right)}\left(\vartheta_{i}=0\right)=\ell_{i}>0$.

\section{\label{sec:Simulation-study}Simulation study}

Levels of gene expression and corresponding observations were simulated
for 2000 gene expression experiments each with $\pi_{0}=90\%$ probability
that any gene is equivalently, $n=2$ observations per gene, and $m=10^{4}$
genes, as follows. For each experiment, the mean differential expression
levels $\theta_{1},\dots,\theta_{m}$ were independently assigned
$0$ with probability $\pi_{0}$, $-2$ with probability $\left(1-\pi_{0}\right)/2$,
and $+2$ with probability $\left(1-\pi_{0}\right)/2$. Then, for
each $i=1,\dots,m$, the $n$ observed expression levels were independently
drawn from $\N\left(\theta_{i},\sigma_{i}^{2}\right)$, where $\sigma_{i}=1$
if $\theta_{i}=0$ and $\sigma_{i}=3/2$ if $\theta_{i}\ne0$, in
accordance with Examples \ref{exa:paired-1gene} and \ref{exa:paired-multiple-genes}.
The posterior medians $\hat{F}_{x_{1}}^{-1}\left(50\%\right)$ and
$\hat{F}_{\left(1\right)}^{-1}\left(50\%\right)$ and the 95\% confidence
intervals $\left[F_{x_{1}}^{-1}\left(2.5\%\right),F_{x_{1}}^{-1}\left(97.5\%\right)\right]$
and $\left[\hat{F}_{\left(1\right)}^{-1}\left(2.5\%\right),\hat{F}_{\left(1\right)}^{-1}\left(97.5\%\right)\right]$
were computed for each simulated experiment. A total of $2000$ experiments
were thereby simulated and analyzed. To assess dependence on the proportion
of true null hypotheses, all of the simulations and analyses were
repeated for $\pi_{0}=99\%$ using the same seed of the pseudo-random
numbers.

\begin{figure}[ph]
\includegraphics{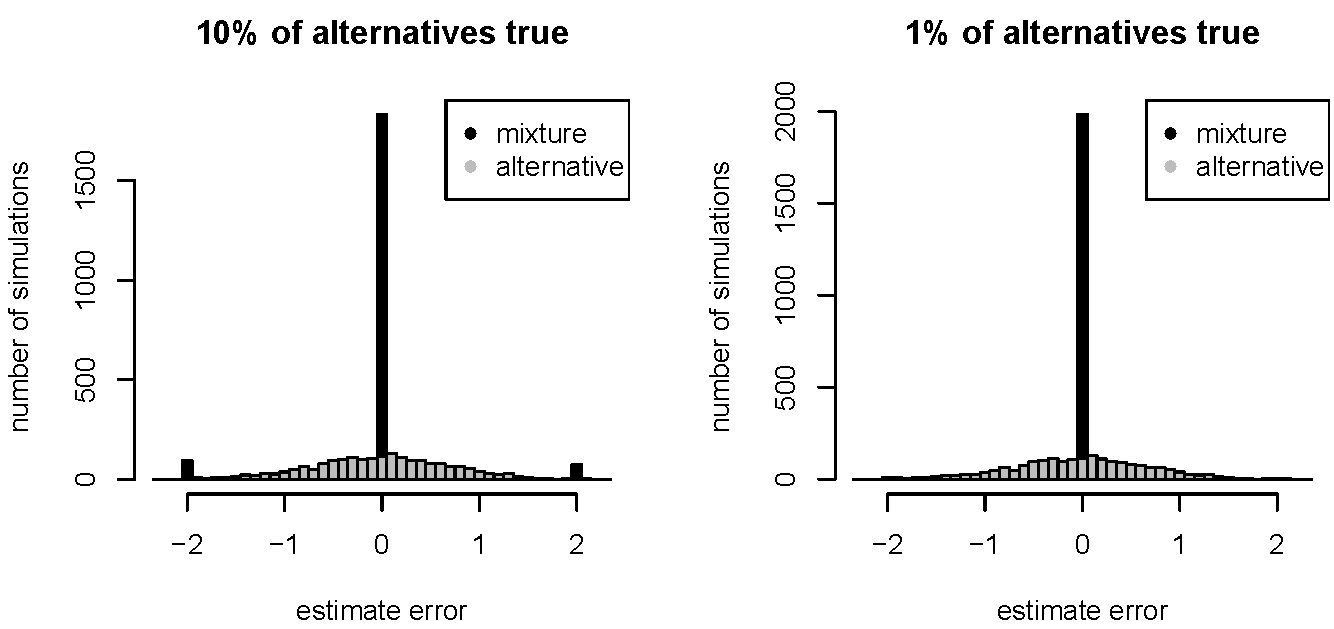}

\caption{Point estimate performance.\label{fig:point-estimate-accuracy}}

\end{figure}

Fig. \ref{fig:point-estimate-accuracy} consists of histograms of
the posterior median errors $\hat{F}_{\left(i\right)}^{-1}\left(1/2\right)-\theta_{i}$
(black) and $F_{x_{i}}^{-1}\left(1/2\right)-\theta_{i}$ (gray) according
to each marginal confidence posterior $\hat{P}^{\left(i\right)}$
and each conditional confidence posterior $P^{x_{i}}$, respectively.
The left panel corresponds to $1-\pi_{0}=10\%$ and the right panel
to $1-\pi_{0}=1\%$.

\begin{figure}[ph]
\includegraphics{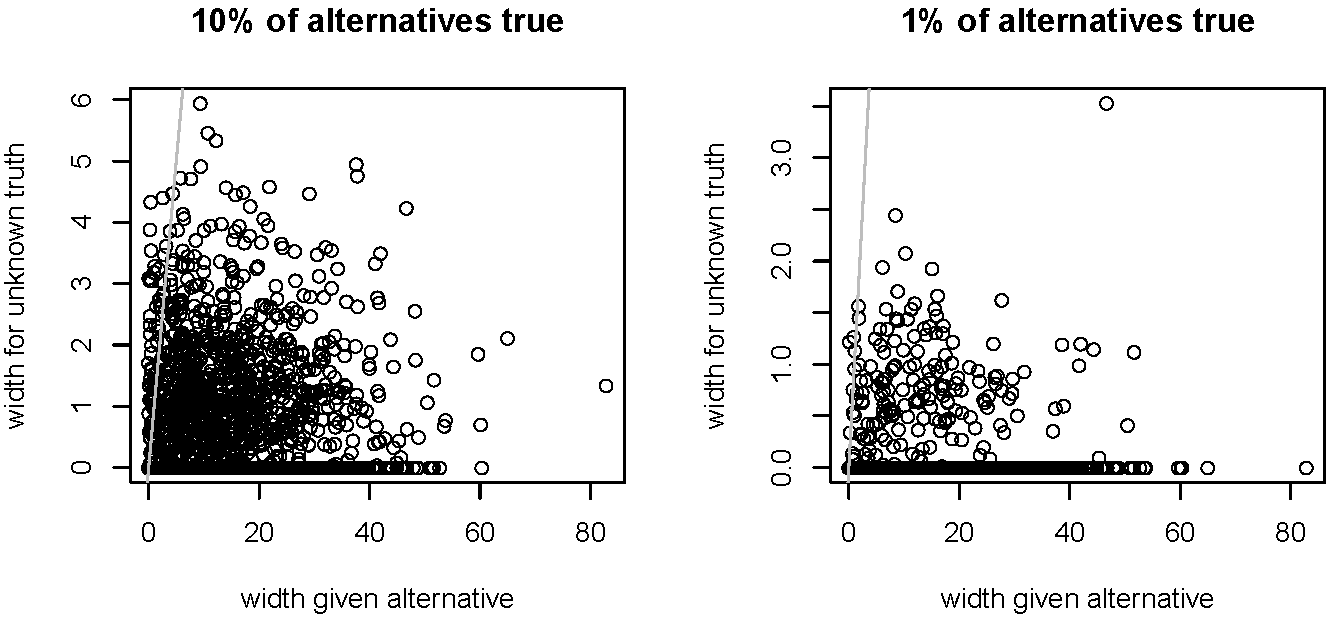}

\caption{95\% confidence interval performance.\label{fig:interval-estimate-performance}}

\end{figure}

Fig. \ref{fig:interval-estimate-performance} gives the width of the
confidence interval from each $\hat{P}^{\left(i\right)}$ versus the
width of the confidence interval from each $P^{x_{i}}$ for $1-\pi_{0}=10\%$
(left) and $1-\pi_{0}=1\%$ (right). Each circle corresponds to a
simulated experiment. As seen in the application to gene expression
(§\ref{sec:Case-study}), the marginal confidence intervals tend to
be much shorter than the conditional confidence intervals. 

The smaller intervals do not compromise frequentist coverage. On the
contrary, the confidence intervals from $\hat{P}^{\left(i\right)}$
cover the simulated values of $\theta_{i}$ at rates higher than the
nominal 95\% level (Table \ref{tab:fig:interval-estimate-coverage}),
in agreement with equations \eqref{eq:approaching0} and \eqref{eq:approaching1}.

\section{\label{sec:Discussion}Discussion}

As an extension of both a confidence posterior and an empirical Bayes
posterior, $\hat{P}^{\left(i\right)}$ offers new approaches to two
related problems in high-dimensional biology. First, the problem of
prioritizing biological features for further study was addressed by
ranking the features according to their $\hat{P}^{\left(i\right)}$-posterior
means or medians. Since a point estimate for an individual feature
of scientific interest is difficult to interpret without an indication
of its reliability, the problem of reporting interval estimates consonant
with the point estimates was handled by constructing the confidence
intervals of each feature to have the posterior median at its center.
The next two paragraphs summarize the findings relevant to each proposed
solution in turn.

The posterior median of $\hat{P}^{\left(i\right)}$ is suitable for
ranking features in order of priority or interest since it is parameterization-invariant
and since it adjusts the uncorrected parameter estimate according
to statistical significance as recorded in the LFDR. The commonly
used alternative of using the LFDR or other measure of significance
to make and accept-reject decision followed by conventional estimation
of the parameter does not perform well since it depends on an arbitrary
threshold to distinguish acceptance from rejection \citep{shrink2009}.
The simulations show that the posterior median of $\hat{P}^{\left(i\right)}$
does perform well in terms of hitting or coming close to its target
parameter value (Fig. \ref{fig:point-estimate-accuracy}).

The confidence intervals based on $\hat{P}^{\left(i\right)}$ are
not only centered at the estimates recommended for ranking features,
but also tend to be much shorter than the fixed-parameter confidence
intervals on which they are based, as seen both in the application
to gene expression (Fig. \ref{fig:expression}) and in the simulation
study (Fig. \ref{fig:interval-estimate-performance}). In spite of
their shortness, the shrunken confidence intervals cover their target
parameter values at rates higher than those claimed (Table \ref{tab:fig:interval-estimate-coverage}).

Some caution is needed in interpreting $\hat{P}^{\left(i\right)}\left(\vartheta_{i}<\theta_{0}\right)$,
$\hat{P}^{\left(i\right)}\left(\vartheta_{i}=\theta_{0}\right)$,
$\hat{P}^{\left(i\right)}\left(\vartheta_{i}>\theta_{0}\right)$,
and other observed marginal confidence levels as posterior probabilities
for decision-making purposes. Since the LFDR estimate $\hat{\ell}_{i}$
is conservative in the sense that it has an upward bias \citep{Pawitan20053865,GWAselect},
the $\hat{P}^{\left(i\right)}$-probability of any hypothesis that
includes or excludes $\theta_{0}$ will tend to be too high or too
low, respectively. For example, there is no warrant for concluding
from $\hat{P}^{\left(i\right)}\left(\vartheta_{i}=\theta_{0}\right)=\hat{\ell}_{i}=100\%$
that the null hypothesis is true with absolute certainty \citep{conditional2009,GWAselect},
and the observed conditional confidence of Fig. \ref{fig:Observed-confidence-levels}
and studied by \citet{conditional2009} would thus perform better
in terms of logarithmic loss or other scoring rules that infinitely
penalize predicting an event with certainty that does not occur.

\section*{Acknowledgments}

The \texttt{Biobase} \citep{RefWorks:161} and \texttt{locfdr} \citep{RefWorks:208}
packages of \texttt{R} \citep{R2008} facilitated the computational
work. This research was partially supported  by the Canada Foundation
for Innovation, by the Ministry of Research and Innovation of Ontario,
and by the Faculty of Medicine of the University of Ottawa. 

The author thanks Xuemei Tang for providing the fruit development
microarray data.

\begin{flushleft}
\bibliographystyle{elsarticle-harv}
\bibliography{refman}

\begin{thebibliography}{35}
\expandafter\ifx\csname natexlab\endcsname\relax\def\natexlab#1{#1}\fi
\expandafter\ifx\csname url\endcsname\relax
  \def\url#1{\texttt{#1}}\fi
\expandafter\ifx\csname urlprefix\endcsname\relax\def\urlprefix{URL }\fi

\bibitem[{Alba et~al.(2005)Alba, Payton, Fei, McQuinn, Debbie, Martin,
  Tanksley, and Giovannoni}]{RefWorks:8}
Alba, R., Payton, P., Fei, Z., McQuinn, R., Debbie, P., Martin, G.~B.,
  Tanksley, S.~D., Giovannoni, J.~J., 2005. Transcriptome and selected
  metabolite analyses reveal multiple points of ethylene control during tomato
  fruit development. Plant Cell 17~(11), 2954--2965.

\bibitem[{Benjamini and Hochberg(1995)}]{RefWorks:288}
Benjamini, Y., Hochberg, Y., 1995. Controlling the false discovery rate: A
  practical and powerful approach to multiple testing. Journal of the Royal
  Statistical Society B 57, 289--300.

\bibitem[{Benjamini et~al.(2005)Benjamini, Yekutieli, Edwards, Shaffer,
  Tamhane, Westfall, Holland, Benjamini, and Yekutieli}]{RefWorks:1276}
Benjamini, Y., Yekutieli, D., Edwards, D., Shaffer, J.~P., Tamhane, A.~C.,
  Westfall, P.~H., Holland, B., Benjamini, Y., Yekutieli, D., 2005. False
  discovery rate-adjusted multiple confidence intervals for selected
  parameters. Journal of the American Statistical Association 100~(469),
  71--93.

\bibitem[{Berger and Pericchi(1996)}]{RefWorks:1023}
Berger, J.~O., Pericchi, L.~R., 1996. The intrinsic {B}ayes factor for model
  selection and prediction. Journal of the American Statistical Association
  91~(433), 109--122.

\bibitem[{Bickel(2004)}]{antiFDRcontrol2004}
Bickel, D.~R., 2004. On 'strong control, conservative point estimation and
  simultaneous conservative consistency of false discovery rates': Does a large
  number of tests obviate confidence intervals of the fdr? Technical Report,
  Office of Biostatistics and Bioinformatics, Medical College of Georgia,
  arXiv:q-bio/0404032.

\bibitem[{Bickel(2008)}]{RefWorks:26}
Bickel, D.~R., 2008. Correcting the estimated level of differential expression
  for gene selection bias: Application to a microarray study. Statistical
  Applications in Genetics and Molecular Biology 7~(1), 10.

\bibitem[{Bickel(2010{\natexlab{a}})}]{CoherentFrequentism}
Bickel, D.~R., 2010{\natexlab{a}}. Coherent frequentism: A decision theory
  based on confidence sets. to appear in Communications in Statistics - Theory
  and Methods, preprint at arXiv:0907.0139.

\bibitem[{Bickel(2010{\natexlab{b}})}]{conditional2009}
Bickel, D.~R., 2010{\natexlab{b}}. Estimating the null distribution to adjust
  observed confidence levels for genome-scale screening. Biometrics, DOI:
  10.1111/j.1541-0420.2010.01491.x.

\bibitem[{Carlin and Louis(2009)}]{CarlinLouis3}
Carlin, B.~P., Louis, T.~A., 2009. {B}ayesian Methods for Data Analysis, Third
  Edition. Chapman \& Hall/CRC, New York.

\bibitem[{Efron(1993)}]{Efron19933}
Efron, B., 1993. {B}ayes and likelihood calculations from confidence intervals.
  Biometrika 80, 3--26.

\bibitem[{Efron(2004)}]{RefWorks:55}
Efron, B., 2004. Large-scale simultaneous hypothesis testing: The choice of a
  null hypothesis. Journal of the American Statistical Association 99~(465),
  96--104.

\bibitem[{Efron(2007)}]{RefWorks:208}
Efron, B., 2007. Size, power and false discovery rates. Annals of Statistics
  35, 1351--1377.

\bibitem[{Efron(2008)}]{RefWorks:1275}
Efron, B., 2008. Microarrays, empirical {B}ayes and the two-groups model.
  Statistical Science 23~(1), 1--22.

\bibitem[{Efron(2010)}]{efron_correlated_2010}
Efron, B., Sep. 2010. Correlated z-values and the accuracy of large-scale
  statistical estimates. Journal of the American Statistical Association
  105~(491), 1042--1055, {PMID:} 21052523.

\bibitem[{Efron and Tibshirani(2002)}]{RefWorks:54}
Efron, B., Tibshirani, R., 2002. Empirical {B}ayes methods and false discovery
  rates for microarrays. Genetic epidemiology 23~(1), 70--86.

\bibitem[{Efron et~al.(2001)Efron, Tibshirani, Storey, and
  Tusher}]{RefWorks:53}
Efron, B., Tibshirani, R., Storey, J.~D., Tusher, V., 2001. Empirical {B}ayes
  analysis of a microarray experiment. J. Am. Stat. Assoc. 96~(456),
  1151--1160.

\bibitem[{Farcomeni(2008)}]{Farcomeni2008347}
Farcomeni, A., 2008. A review of modern multiple hypothesis testing, with
  particular attention to the false discovery proportion. Statistical Methods
  in Medical Research 17~(4), 347--388.

\bibitem[{Fraser(1991)}]{RefWorks:60}
Fraser, D. A.~S., 1991. Statistical inference: likelihood to significance.
  Journal of the American Statistical Association 86, 258--265.

\bibitem[{Gentleman et~al.(2004)Gentleman, Carey, Bates, et~al.}]{RefWorks:161}
Gentleman, R.~C., Carey, V.~J., Bates, D.~M., et~al., 2004. Bioconductor: Open
  software development for computational biology and bioinformatics. Genome
  Biology 5, R80.

\bibitem[{Ghosh(2009)}]{RefWorks:1273}
Ghosh, D., 2009. Empirical {B}ayes methods for estimation and confidence
  intervals in high-dimensional problems. Statistica Sinica 19~(1), 125--143.

\bibitem[{Montazeri et~al.(2010)Montazeri, Yanofsky, and Bickel}]{shrink2009}
Montazeri, Z., Yanofsky, C.~M., Bickel, D.~R., 2010. Shrinkage estimation of
  effect sizes as an alternative to hypothesis testing followed by estimation
  in high-dimensional biology: Applications to differential gene expression.
  Statistical Applications in Genetics and Molecular Biology 9, 23.

\bibitem[{Pawitan et~al.(2005)Pawitan, Murthy, Michiels, and
  Ploner}]{Pawitan20053865}
Pawitan, Y., Murthy, K., Michiels, S., Ploner, A., 2005. Bias in the estimation
  of false discovery rate in microarray studies. Bioinformatics 21~(20),
  3865--3872.

\bibitem[{Polansky(2007)}]{Polansky2007b}
Polansky, A.~M., 2007. Observed Confidence Levels: Theory and Application.
  Chapman and Hall, New York.

\bibitem[{Qiu et~al.(2005)Qiu, Klebanov, and Yakovlev}]{Qiu2005i}
Qiu, X., Klebanov, L., Yakovlev, A., 2005. Correlation between gene expression
  levels and limitations of the empirical {B}ayes methodology for finding
  differentially expressed genes. Statistical Applications in Genetics and
  Molecular Biology 4~(1), i--30.

\bibitem[{{R Development Core Team}(2008)}]{R2008}
{R Development Core Team}, 2008. R: A language and environment for statistical
  computing. R Foundation for Statistical Computing, Vienna, Austria.

\bibitem[{Savage(1954)}]{RefWorks:126}
Savage, L.~J., 1954. The Foundations of Statistics. John Wiley and Sons, New
  York.

\bibitem[{Schwartzman et~al.(2009)Schwartzman, Dougherty, Lee, Ghahremani, and
  Taylor}]{Schwartzman200971}
Schwartzman, A., Dougherty, R.~F., Lee, J., Ghahremani, D., Taylor, J.~E.,
  2009. Empirical null and false discovery rate analysis in neuroimaging.
  NeuroImage 44~(1), 71 -- 82.

\bibitem[{Schweder and Hjort(2002)}]{RefWorks:127}
Schweder, T., Hjort, N.~L., 2002. Confidence and likelihood. Scandinavian
  Journal of Statistics 29~(2), 309--332.

\bibitem[{Singh et~al.(2005)Singh, Xie, and Strawderman}]{RefWorks:130}
Singh, K., Xie, M., Strawderman, W.~E., 2005. Combining information from
  independent sources through confidence distributions. Annals of Statistics
  33~(1), 159--183.

\bibitem[{Stromberg et~al.({2008})Stromberg, Bjork, Broberg, Mertens, and
  Vineis}]{ISI:000253401400024}
Stromberg, U., Bjork, J., Broberg, K., Mertens, F., Vineis, P., {MAR} {2008}.
  {Selection of influential genetic markers among a large number of candidates
  based on effect estimation rather than hypothesis testing - {A}n approach for
  genome-wide association studies}. {EPIDEMIOLOGY} {19}~({2}), {302--308}.

\bibitem[{von Neumann and Morgenstern(1944)}]{MaxUtility1944}
von Neumann, J., Morgenstern, O., 1944. Theory of Games and Economic Behavior.
  Princeton University Press, Princeton.

\bibitem[{Wei et~al.({2010})Wei, Wen, Chen, Wang, and
  Hsiao}]{ISI:000280145100013}
Wei, Y.-C., Wen, S.-H., Chen, P.-C., Wang, C.-H., Hsiao, C.~K., {AUG} {2010}.
  {A simple {B}ayesian mixture model with a hybrid procedure for genome-wide
  association studies}. {EUROPEAN JOURNAL OF HUMAN GENETICS} {18}~({8}),
  {942--947}.

\bibitem[{Westfall(2010)}]{Westfall_efron_correlated_2010}
Westfall, P.~H., Sep. 2010. Comment on b. efron, "correlated z-values and the
  accuracy of large-scale statistical estimates". Journal of the American
  Statistical Association 105~(491), 1063--1066, {PMID:} 21052523.

\bibitem[{Yang and Bickel(2010)}]{GWAselect}
Yang, Y., Bickel, D.~R., 2010. Minimum description length and empirical bayes
  methods of identifying snps associated with disease. Technical Report, Ottawa
  Institute of Systems Biology, COBRA Preprint Series, Article 74, available at
  biostats.bepress.com/cobra/ps/art74.

\bibitem[{Yanofsky and Bickel(2010)}]{mse2008}
Yanofsky, C.~M., Bickel, D.~R., 2010. Validation of differential gene
  expression algorithms: Application comparing fold-change estimation to
  hypothesis testing. BMC Bioinformatics 11, 63.

\end{thebibliography}

\par\end{flushleft}

\newpage{}


\newpage{}%
\begin{table}[ph]
\begin{tabular}{ccc}
\hline 
 & $1-\pi_{0}=10\%$ & $1-\pi_{0}=1\%$\tabularnewline
\hline
\hline 
Marginal confidence & 97.5\% & 99.2\%\tabularnewline
\hline 
Conditional confidence & 95.3\% & 95.3\%\tabularnewline
\hline
\end{tabular}

\caption{95\% confidence interval coverage.\label{tab:fig:interval-estimate-coverage}}

\end{table}

\end{document}